\documentclass[%
 reprint,
preprintnumbers,
nofootinbib,
 amsmath,amssymb,
 aps,
float,
superscriptaddress,
showkeys
]{revtex4-2}

\usepackage{graphicx}

\usepackage{bm}
\usepackage[mathlines]{lineno}
\usepackage[table]{xcolor}
\usepackage{multirow}
\usepackage[title]{appendix}

\begin{document}

\preprint{APS/123-QED}
\title{$^{22}$Na Activation Level Measurements of Fused Silica Rods in the LHC Target Absorber for Neutrals (TAN) Compared to FLUKA Simulations}

\author{S.~Yang}
\affiliation{Department of Physics, University of Illinois, 1110 W. Green St., Urbana IL 61801-3080, USA}
\affiliation{Department of Nuclear, Plasma and Radiological Engineering, University of Illinois, 1110 W. Green St., Urbana IL 61801-3080, USA}

\author{M.~Sabate~Gilarte} 
\affiliation{CERN, CH-1211 Geneva 23, Switzerland}

\author{A.~Tate}
\affiliation{Department of Physics, University of Illinois, 1110 W. Green St., Urbana IL 61801-3080, USA}
\affiliation{Department of Nuclear, Plasma and Radiological Engineering, University of Illinois, 1110 W. Green St., Urbana IL 61801-3080, USA}

\author{N.~Santiago}
\affiliation{Department of Physics, University of Illinois, 1110 W. Green St., Urbana IL 61801-3080, USA}
\affiliation{Nuclear Engineering Division, Argonne National Laboratory, Lemont, IL, USA}

\author{R.~Longo}
\affiliation{Department of Physics, University of Illinois, 1110 W. Green St., Urbana IL 61801-3080, USA}

\author{S.~Mazzoni}
\affiliation{CERN, CH-1211 Geneva 23, Switzerland}

\author{F.~Cerutti}
\affiliation{CERN, CH-1211 Geneva 23, Switzerland}

\author{E.~Bravin}
\affiliation{CERN, CH-1211 Geneva 23, Switzerland}

\author{M.~Grosse~Perdekamp}
\affiliation{Department of Physics, University of Illinois, 1110 W. Green St., Urbana IL 61801-3080, USA}

\author{G.~Lerner}
\affiliation{CERN, CH-1211 Geneva 23, Switzerland}

\author{D.~Prelipcean}
\affiliation{CERN, CH-1211 Geneva 23, Switzerland}
\affiliation{Technical University of Munich (TUM), Department of Physics, Munich, Germany}

\author{Z.~Citron}
\affiliation{Ben-Gurion University of the Negev, Dept.\ of Physics, Beer-Sheva 84105, Israel}

\author{B.~Cole}
\affiliation{Columbia University, New  York,  New  York  10027  and  Nevis  Laboratories,  Irvington,  New  York  10533,  USA }

\author{ S.~Jackobsen}
\affiliation{CERN, CH-1211 Geneva 23, Switzerland}

\author{M.~D.~Kaminski} 
\affiliation{Department of Nuclear, Plasma and Radiological Engineering, University of Illinois, 1110 W. Green St., Urbana IL 61801-3080, USA}
\affiliation{Nuclear Engineering Division, Argonne National Laboratory, Lemont, IL, USA}

\author{T.Koeth}
\affiliation{University of Maryland, Dept.\ of Chemistry and Biochemistry, College Park, MD 20742, USA}

\author{C.~Lantz} 
\affiliation{Department of Physics, University of Illinois, 1110 W. Green St., Urbana IL 61801-3080, USA}

\author{D.~MacLean}
\affiliation{Department of Physics, University of Illinois, 1110 W. Green St., Urbana IL 61801-3080, USA}

\author{A.~Mignerey}
\affiliation{University of Maryland, Dept.\ of Chemistry and Biochemistry, College Park, MD 20742, USA}

\author{M.~Murray}
\affiliation{University of Kansas, Dept.\ of Physics, Lawrence, KS 66045, USA}

\author{M.~Palm}
\affiliation{CERN, CH-1211 Geneva 23, Switzerland}

\author{M.~Phipps}
\affiliation{Department of Physics, University of Illinois, 1110 W. Green St., Urbana IL 61801-3080, USA}

\author{P.~Steinberg}
\affiliation{Brookhaven National Laboratory, Upton, NY 11973, USA}

\author{A.~Tsinganis}
\affiliation{CERN, CH-1211 Geneva 23, Switzerland}
\affiliation{nowadays with European Commission, Joint Research Centre (JRC), Geel, Belgium}

\date{\today}
\begin{abstract}
The Target Absorbers for Neutrals (TANs) are located in a high intensity radiation environment inside the tunnel of the Large Hadron Collider (LHC). TANs are positioned about $140$\,m downstream from the beam interaction points. Seven $40$\,cm long fused silica rods with different dopant specifications were irradiated in the TAN by the Beam RAte of Neutrals (BRAN) detector group during $p$+$p$  data taking from 2016 to 2018 at the LHC. The peak dose delivered to the fused silica rods was $18$\,MGy. 
We report measurements of the $^{22}$Na activation of the fused silica rods carried out at the University of Illinois at Urbana-Champaign and Argonne National Laboratory. At the end of the irradiation campaign, the maximum $^{22}$Na activity observed was $A=21$\,kBq$/{\rm cm^3}$ corresponding to a density, $\rho= 2.5\times 10^{12} /{\rm cm^3}$, of $^{22}$Na nuclei.
FLUKA Monte Carlo simulations have been performed by the CERN FLUKA team to estimate $^{22}$Na activities for the irradiated BRAN rod samples.  
The simulations reproduce the $^{22}$Na activity profile measured along the rods, with a 35\% underestimation of the experimental measurement results. 
\end{abstract}

\maketitle
\section{Introduction}
\label{sec:introduction}
The High Luminosity upgrade of the Large Hadron Collider (HL-LHC) places significant demands on the radiation hardness of the detectors installed in the forward region inside the Large Hadron Collider (LHC) tunnel. These detectors are integrated inside the upgraded Target Absorber for 
Neutrals (TAXN) and comprise the Beam RAte of Neutrals (BRAN)~\cite{Matis:2016raz} and Zero Degree Calorimeter (ZDC). The radiation levels in the TAXN throughout Run 4 (first years of HL-LHC operation) are expected to be increased by a factor of 3 to 4 compared to the present TAN 
throughout Run 2 (LHC operation from 2015 to 2018). As a consequence, forward detector groups have begun R\&D to identify radiation-hard materials to design and construct the new generation of detectors for the high luminosity era.  

Fused silica is a promising radiation hard medium with strong light transmission properties, for example for Cherenkov detectors \cite{hoek2008radiation, hoek2011tailoring} and scintillation detectors \cite{akchurin2018cerium}. During the 2016-2018 $p$+$p$  running, an additional radiation-hard BRAN detector prototype, based on fused silica rods, was installed in the TAN. 
The rods in the BRAN prototype were irradiated in order to study the degradation of the light yield as a function of delivered luminosity and corresponding dose. The latter was estimated from simulations performed by the CERN FLUKA team. FLUKA \cite{FLUKA:web, FLUKA:new, FLUKA:old} is a simulation framework used in many applications, and in particular studies related to radiation damage and radiation protection  \cite{skordisPhD,brugger2006validation,rata2016fluka,iliopoulou2018measurements}. 

The unprecedented operational energy reached by the LHC during Run 2 gave us the opportunity to investigate FLUKA's performance in describing activation of the materials irradiated in the LHC. 
To perform such studies, dedicated FLUKA simulations were carried out to estimate the activation of the BRAN fused silica samples. In this paper, the simulation results are compared to the measurements performed independently at the University of Illinois at Urbana-Champaign (UIUC) and Argonne National Laboratory (ANL).

The paper is structured as follows: Sec.~\ref{sec:BRAN} describes the BRAN irradiation setup, Sec.~\ref{sec:FLUKA} introduces the FLUKA simulation model and the dose and activation estimates for the fused silica samples. In Sec.~\ref{sec:Activation}, the details of the activation measurements at UIUC and ANL are described. The data analysis procedures are then discussed in Sec.~\ref{sec:DataAnalysis}.  Results are presented and discussed in Sec.~\ref{sec:Results}. Finally, the conclusions are given in Sec.~\ref{sec:Conclusions}. 

%
%
\begin{figure*}[!htp]
    \centering
    \includegraphics[width=0.75\linewidth,keepaspectratio]{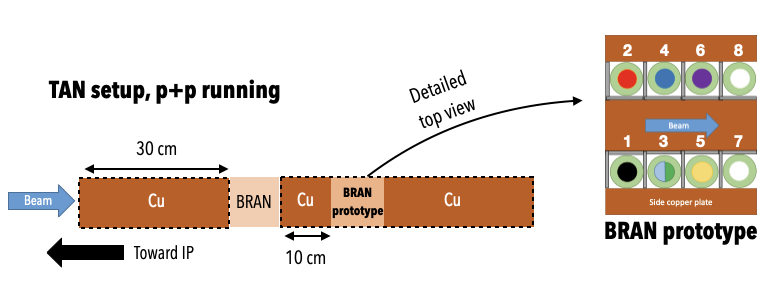}
    \caption{TAN setup during LHC Run 2 $p$+$p$  running (2016-2018). The dashed lines identify the slot used to install the Zero Degree Calorimeter during the Heavy Ion runs. The right insert shows a top view of the BRAN prototype. The numbers identify the position of the fused silica rods. Detailed information on the materials of the rods and maximum irradiation reached for each of them are reported in Tab.~\ref{tab:BRANrods}.}
     \label{fig:BRANSet}
\end{figure*}

\definecolor{mayablue}{rgb}{0.45, 0.76, 0.98}
\definecolor{lapislazuli}{rgb}{0.15, 0.38, 0.61}
\definecolor{meatbrown}{rgb}{0.9, 0.72, 0.23}
\definecolor{purpleheart}{rgb}{0.41, 0.21, 0.61}
\definecolor{pakistangreen}{rgb}{0.0, 0.4, 0.0}
\begin{table*}[!htbp]
\centering
\caption{Fused silica rod specifications. The number ID assigned to each rod corresponds to a given position in the detector, as shown in Fig.~\ref{fig:BRANSet}. The rods were doped with different levels of hydroxyl (OH) and hydrogen (H$_2$) to carry out optical transmission studies ~\cite{Transmission:2022}. Rods 3a and 3b occupied the same slot, but at different times.}
\vspace{0.2cm}
\begin{tabular}{|c|c|c|c|c|c|}
\hline
\textbf{BRAN}       & \textbf{Irradiation}  & \textbf{Max. Exposure}    & \multirow{2}{*}{\textbf{Material}}    & \textbf{H$_{2}$}  & \textbf{OH} \\ 
\textbf{Position}   & \textbf{Period}       & [MGy]                    &                                       & [mol/cm$^{3}$]    & [ppm] \\ 
\hline
\hline 

\multirow{2}{*}{\textbf{Control}}    &  \multirow{2}{*}{None}    & \multirow{2}{*}{0}    & Spectrosil 2000   & \multirow{2}{*}{7.20e17}     & \multirow{2}{*}{1120} \\ 
                            &                           &                       & (High OH, Mid H$_{2}$) &                           &  \\ 
\hline

\cellcolor{black}        &  04/2016 -    & \multirow{2}{*}{18}    & Spectrosil 2000   & \multirow{2}{*}{7.20e17}      & \multirow{2}{*}{1120} \\ 
\multirow{-2}{*}{\cellcolor{black}\textbf{\textcolor{white}1}}   &   12/2018     &                       & (High OH, Mid H$_{2}$) &                               &  \\ 
\hline
\cellcolor{red}                                                 &  04/2016 -    & \multirow{2}{*}{10}      & Spectrosil 2000   & \multirow{2}{*}{7.20e17}      & \multirow{2}{*}{1120} \\ 
\multirow{-2}{*}{\cellcolor{red}\textbf{\textcolor{white}2}}   &   12/2017     &                           & (High OH, Mid H$_{2}$) &                               &  \\ 
\hline
\cellcolor{pakistangreen}                                                   &  04/2016 -    & \multirow{2}{*}{5}      & Spectrosil 2000   & \multirow{2}{*}{2.80e18}      & \multirow{2}{*}{1000} \\ 
\multirow{-2}{*}{\cellcolor{pakistangreen}\textbf{\textcolor{white}{3a}}}    &   12/2016     &                           & (High OH, High H$_{2}$) &                               &  \\ 
\hline
\cellcolor{mayablue}                                                &  04/2017 -    & \multirow{2}{*}{16}      & Spectrosil 2000   & \multirow{2}{*}{7.20e17}      & \multirow{2}{*}{1120} \\ 
\multirow{-2}{*}{\cellcolor{mayablue}\textbf{\textcolor{white}{3b}}} &   12/2018     &                           & (High OH, Mid H$_{2}$) &                               &  \\ 
\hline
\cellcolor{lapislazuli}                                                 &  04/2016 -    & \multirow{2}{*}{9}      & Spectrosil 2000   & \multirow{2}{*}{0}      & \multirow{2}{*}{1011} \\ 
\multirow{-2}{*}{\cellcolor{lapislazuli}\textbf{\textcolor{white}4}}   &   12/2017     &                           & (High OH, H$_{2}$ free) &                               &  \\                     
\hline
\cellcolor{meatbrown}                                               &  04/2016 -    & \multirow{2}{*}{8}      & Suprasil 3301   & \multirow{2}{*}{3.00e18}      & \multirow{2}{*}{15} \\ 
\multirow{-2}{*}{\cellcolor{meatbrown}\textbf{\textcolor{white}5}} &   12/2017     &                           & (Low OH, High H$_{2}$) &                               &  \\                     
\hline
\cellcolor{purpleheart}                                                 &  04/2016 -    & \multirow{2}{*}{17}        & Suprasil 3301   & \multirow{2}{*}{0}      & \multirow{2}{*}{14} \\ 
\multirow{-2}{*}{\cellcolor{purpleheart}\textbf{\textcolor{white}6}}   &   12/2018     &                           & (Low OH, H$_{2}$ free) &                               &  \\                     
\hline
\end{tabular}
\label{tab:BRANrods}
\end{table*}

\section{Irradiation setup}
\label{sec:BRAN}

\begin{figure}[!htbp]
    \includegraphics[width=0.95\linewidth,keepaspectratio]{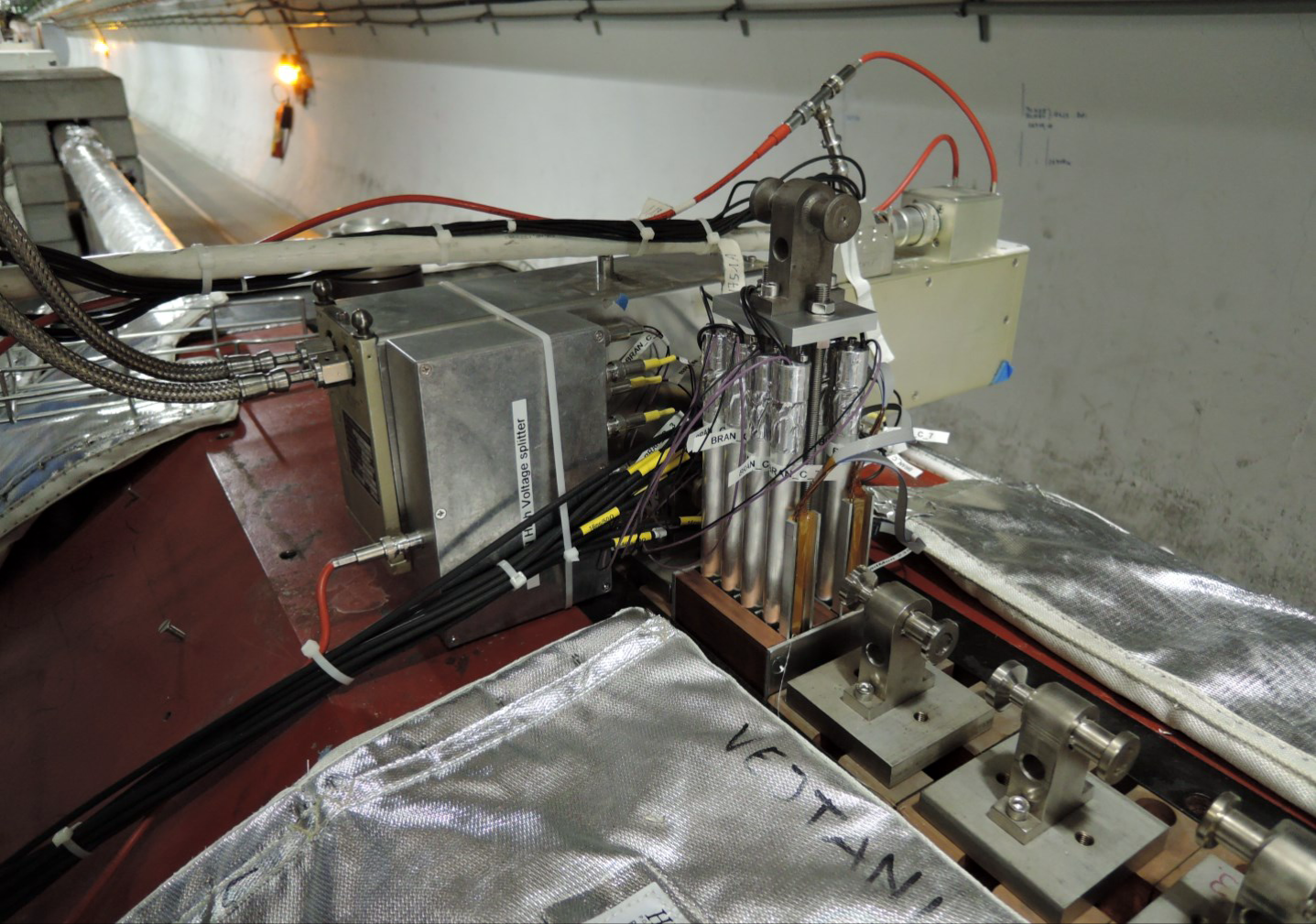}
    \caption{BRAN prototype as installed in 2016 in the TAN on Arm 8-1 of IP1.}
     \label{fig:BRANInstalled}
\end{figure}

The BRAN prototype is a receptacle that can host up to eight silica rod samples. The rods are sandwiched between three copper plates (see Fig.~\ref{fig:BRANSet}) parallel to the beam propagation direction. The prototype can be inserted in the 10 cm - wide gap of the TAN and is equipped with a fixture so that it can be lifted with the dedicated crane installed in the TAN at the ATLAS experiment~\cite{aad2008atlas}. The type of silica rods hosted in positions $1$ to $6$ and produced by Heraeus Quarzglas (Germany) are described in Tab.~\ref{tab:BRANrods}, while $7$ and $8$ were empty and used for testing Cherenkov light yield in air (results not discussed in this paper). To measure the evolution of light transmission during the irradiation process, a photo-multiplier (Hamamatsu R3878P) was positioned on top of each sample holder. The BRAN prototype was first installed in the TAN on Arm 8-1 of IP1 in March 2016, see Fig.~\ref{fig:BRANInstalled}.  

The installed BRAN prototype experienced different beam optics configurations during the irradiation experiment. One of the important parameters is the crossing angle of the LHC beams, which affects the forward distribution of the collision products. When two LHC beams approach each other near an interaction point (IP), they collide with a full angle of a few hundred microradians to prevent other encounters in the region where the two beams share the same vacuum chamber~\cite{evans2008lhc}. 
A small crossing angle increases the overlap area of the bunches, resulting in a higher luminosity. However, the angle has to be large enough to provide a separation that properly contains beam--beam interactions. Moreover, the exploitation of the degree of freedom enabled by vertical crossing  in ATLAS implies a regular polarity inversion (changing the angle sign, with the colliding beams pointing either upward - positive sign - or downward - negative sign) that reduces the peak dose accumulated in the coils of the final focus magnets and so maximizes their lifetime. In fact, the luminosity averaged half crossing angle changed between 2016 and 2017 operation, from -180 $\mu$rad to +140 $\mu$rad.

\section{BRAN dose and activation calculation with FLUKA}
\label{sec:FLUKA}

FLUKA results have shown excellent agreement between simulated and measured values of dose in the complex radiation environment in the LHC~\cite{AntonRun1}. This is achieved thanks to both the quality of the particle interaction and transport implementation, relying on the integration of complementary physics models covering the full range of particle types and energies, and the accuracy of the accelerator geometry description, including material composition and magnetic field maps.  In this paper, we focus on a portion of the ATLAS insertion, from the center (IP1) of the detector hosted in the experimental cavern up to the TAN. The latter is a massive absorber that incorporates the transition between the single vacuum chamber accommodating both beams on one side and two symmetrical apertures of 52 mm diameter connected to separate beam pipes on the opposite side. The copper material in between the two apertures, extending over a 3 m length, fulfils the TAN's protection role, since it intercepts the line of sight of neutral particles emerging from IP1 and thereby shields the downstream superconducting magnets.

\begin{figure}[htbp]
     \centering
     \includegraphics[keepaspectratio,scale=0.23]{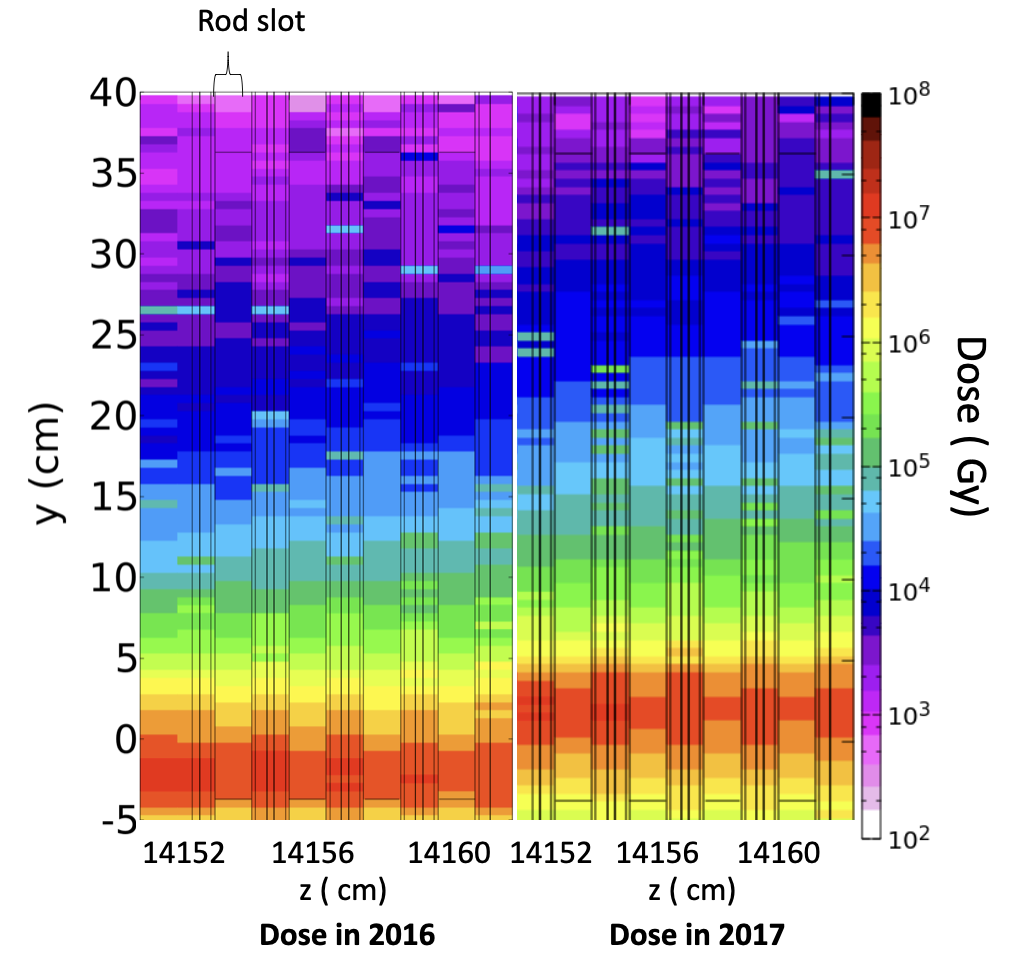}
     \caption{Dose distribution in the BRAN prototype for the 2016 (left) and 2017 (right) $p$+$p$  runs, normalized to an integrated luminosity of 38.5 and 50 fb$^{-1}$, respectively. The horizontal axis gives the distance from IP1, namely the Z-coordinate along the ATLAS detector axis. The vertical axis gives the distance from the beam height, namely the Y-coordinate along the axis opposite to gravity. Dose values are averaged over a 9 mm interval in the missing third dimension, corresponding to the horizontal orthogonal X-axis pointing outside the LHC ring. The vertical shift in the maximum exposure position is due to the crossing angle change.}
     \label{fig:Dosage_profile_YZ}
\end{figure}

\begin{figure*}[ht]
\centering
\includegraphics[width=1\textwidth]{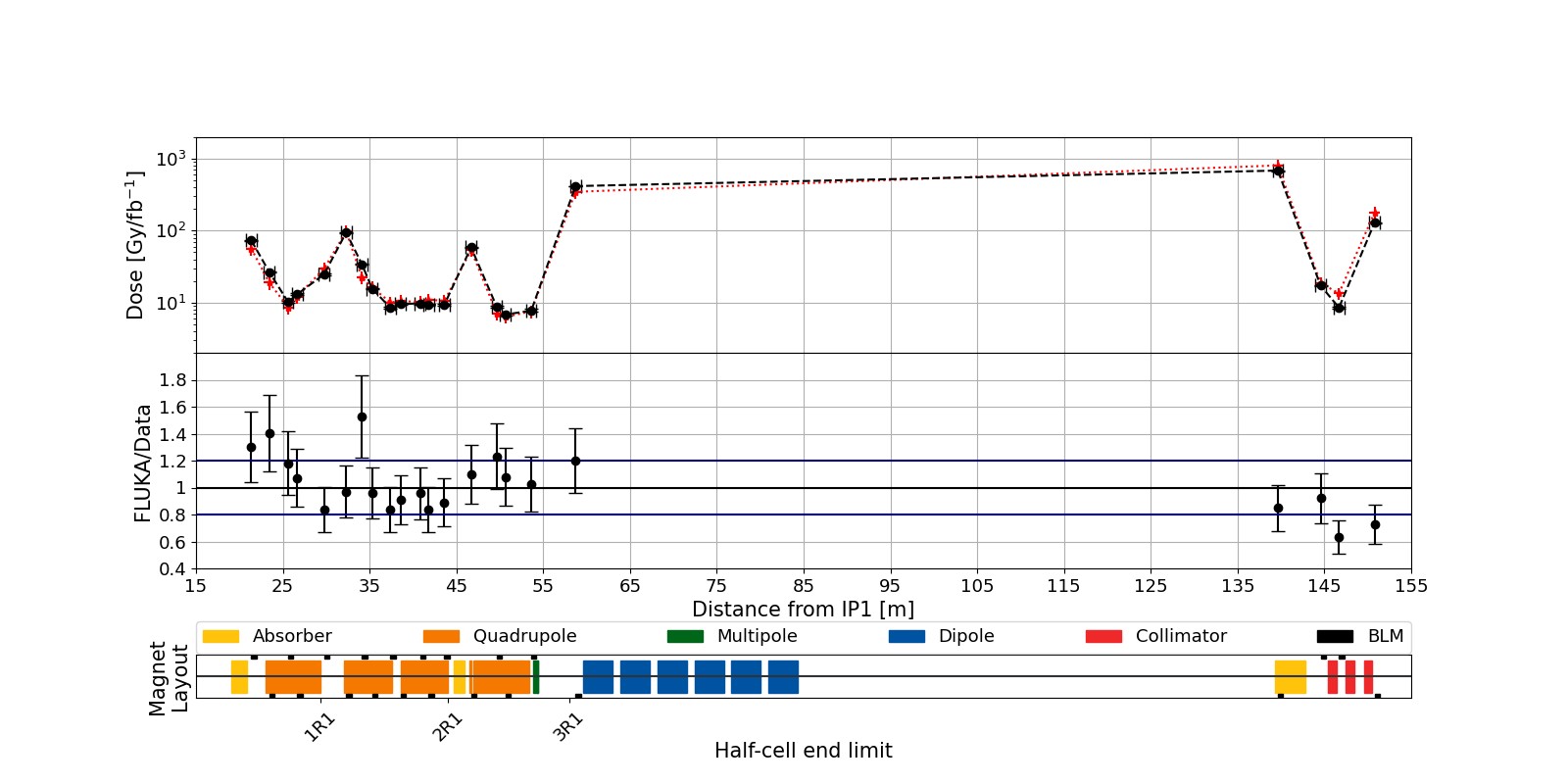}
\caption{Top panel: BLM pattern along the ATLAS insertion as measured over a 2018 13 TeV center-of-mass $p$+$p$  run period yielding 20.3 fb$^{-1}$ (red crosses) and calculated by FLUKA (black circles). Mid panel: Ratio between simulation values and data. Vertical bars correspond to a 20\% uncertainty on data. Bottom panel: Machine layout.}
\label{fig:fluka_sim_exp_comparison}
\end{figure*}

The radiation originates from proton--proton ($p$+$p$) collisions that took place during Run 2 at 13 TeV center-of-mass energy inside ATLAS. Only a small fraction of the collision products reaches the LHC tunnel, through the 34 mm cylindrical aperture of a first 1.8 m long copper absorber called TAS, located at the cavern extremity at about 20 m from IP1. Nevertheless, these are the most energetic particles, carrying 70\% of the collision power onto the machine elements. The forward angle charged component of the collision debris is bent by the strong quadrupole field of the 30 m long string of superconducting magnets ending at a distance of 55 m from IP1.  These magnets, referred to as the triplet, perform the beam squeezing. As a result, they are impacted by the majority of charged pions matching the TAS aperture, which initiate secondary particle showers inside the magnets. These showers deposit into the NbTi cables of the magnets a power density that is safely below the quench threshold but limits their lifetime, due to the peak dose accumulated in the coil insulator. The triplet precedes a normal-conducting single bore separation dipole (D1), consisting of six 3.4 m long modules hosting the two beams in the same pipe, as the previous elements, and extending up to 85 m from IP1, where a 55 m long drift reaching the TAN starts.

As earlier indicated, by design the TAN is impacted by the forward angle neutral component of the collision debris, mainly consisting of photons (from neutral pion immediate decay) and neutrons.
Its protection effectiveness is maximized for the vertical beam crossing in IP1, since in this case the axis of the debris cone hits exactly in between the two TAN twin apertures, while for horizontal crossing it moves closer to the external aperture, as a function of the crossing angle.
In the vertical crossing case which is considered here, the TAN benefits from the significantly increased shadow of the separation dipole, whose geometrical aperture is much smaller in the vertical plane.

Fig.~\ref{fig:Dosage_profile_YZ} shows the distribution of the dose deposited in the BRAN as calculated for the 2016 and 2017 proton--proton runs. One can observe the effect of the inversion of the crossing angle polarity between 2016 and 2017. The maximum dose for each rod can be found in Tab.~\ref{tab:BRANrods}.

Beam Loss Monitor (BLM) measurements offer independent data to benchmark FLUKA simulations. The LHC BLM \cite{zamantzas2006fpga, dehning2007lhc} is an ionization chamber providing an on-line record of the dose deposited in its nitrogen gas by the particle shower originating from beam losses (i.e., collision debris in our case), with multiple time resolutions. As the dose value exceeds a respective pre-defined threshold, beam abort is triggered. A few thousand BLMs are placed all along the 27 km beam line, typically outside the cold magnet cryostats or in the vicinity of collimators.
Fig.~\ref{fig:fluka_sim_exp_comparison} shows the comparison between FLUKA predictions and 2018 data in our region of interest, highlighting an excellent agreement, as a result of a comprehensive simulation accounting for particle transport and re-interaction from IP1 through the whole geometry model that includes a detailed description of the BLMs, allowing for the direct evaluation of the energy released to the gas region. 

\begin{figure}[htbp]
     \centering
     \includegraphics[width=0.45\textwidth]{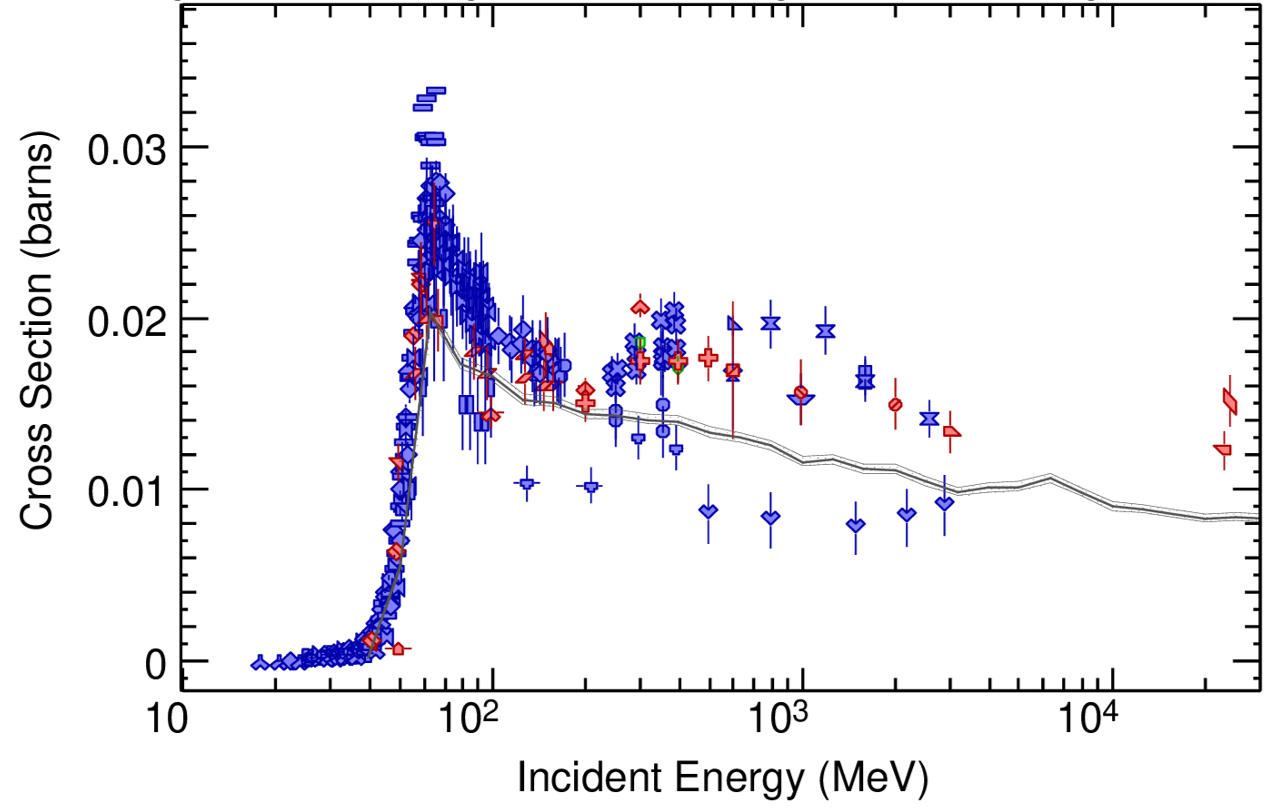}
     \caption{$^{nat}$Si(p,$\star$)$^{22}$Na cross section. Symbols are different experimental data sets from EXFOR \cite{EXFOR}, the grey curve results from the FLUKA interaction model.}
     \label{fig:22Nabyp}
\end{figure}

\begin{figure}[htbp]
     \centering
     \includegraphics[width=0.45\textwidth]{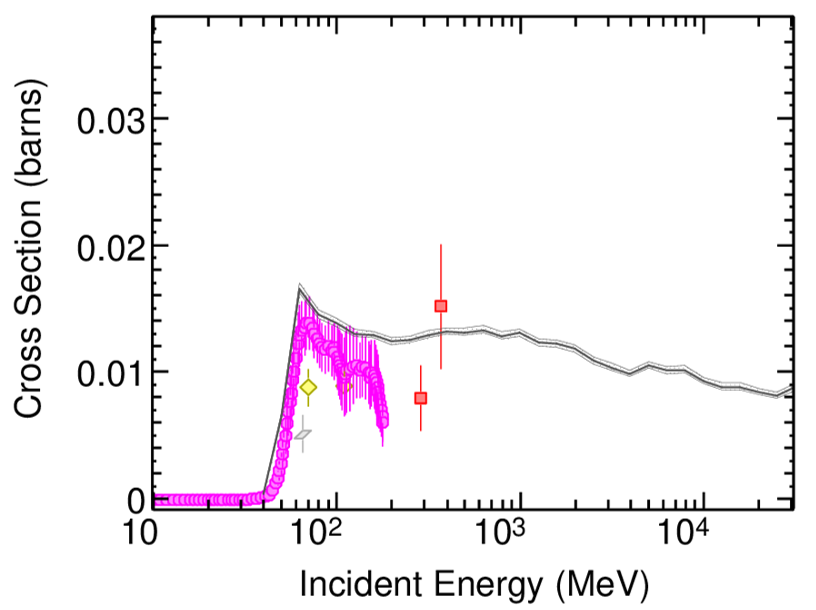}
     \caption{$^{nat}$Si(n,$\star$)$^{22}$Na cross section. Symbols are different experimental data sets from EXFOR \cite{EXFOR}, the grey curve results from the FLUKA interaction model.}
     \label{fig:22Nabyn}
\end{figure}

Nuclear reactions have a major role in shower development, sharing the projectile energy among higher generation particles and feeding the electromagnetic component (that dominates the energy deposition process) through neutral pion production and almost immediate decay. In addition, they are responsible for material activation, with the transformation of the target nucleus into radioactive residue, at the end of the final de-excitation stage. In FLUKA, this stage features
evaporation, fission, gamma emission and/or Fermi breakup, except for low-energy neutron interactions (below 20 MeV) where library data are used. 

For the purposes of our study, a key ingredient is the $^{22}$Na production cross section on silicon by the different particle species travelling through the TAN. Figs.~\ref{fig:22Nabyp} and \ref{fig:22Nabyn} show the cross section over a large energy range for proton and neutron induced reactions, respectively. In addition, a relevant contribution comes from charged pions, while the abundance of photons has a little weight due to the much lower photonuclear cross section.



Two dedicated collections of Monte Carlo results were produced with the following settings: 
\begin{enumerate}
    \item $p$+$p$  - 2016 optics (-180 $\mu$rad crossing angle) and integrated luminosity (38.5 fb$^{-1}$). 
    \item $p$+$p$  - 2017 optics (+140 $\mu$rad crossing angle) and integrated luminosity (50 fb$^{-1}$). 
\end{enumerate}

Each simulation was characterized by different beam crossing angles and irradiation times, depending on the year of LHC data taking. It is worth noting that, apart from the evaluation of prompt dose, FLUKA allows for the on-line calculation of the delayed activity of any produced radionuclide at user defined cooling times, as a function of the input irradiation profile (specifying irradiation intervals and respective beam currents or collision rates). During 2016 and 2017 LHC runs, the rods were irradiated for 198 and 173 days, respectively. In each simulation, the activity was computed at the end of the irradiation period.

Since the 2017 and 2018 runs were characterized by similar crossing angles, it was possible to use the 2017 simulation to estimate the 2018 activation. This step was accomplished by rescaling the results with the ratio of the integrated luminosity delivered by the LHC to ATLAS in the two years and by applying to the results a dedicated cooling correction (Sec.~\ref{sec:cooling_correction}). The activity of each rod after the end of Run 2 was obtained by adding indipendent contributions from different years of LHC running corresponding to the irradiation period in Tab.~\ref{tab:BRANrods}. 

\begin{figure}[htbp]
\includegraphics[width=0.5\textwidth]{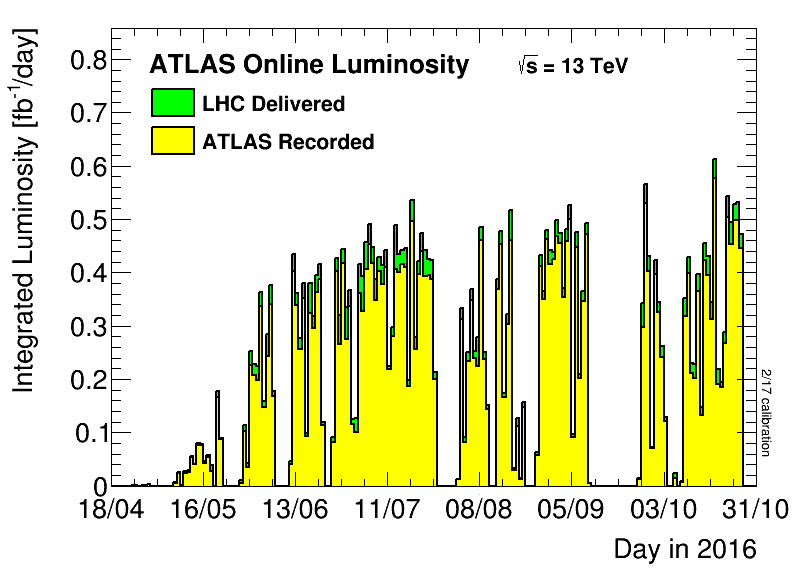}
\caption{Integrated luminosity per day delivered to (green) and recorded by (yellow) ATLAS during stable beams for $p$+$p$  collisions at 13 TeV center-of-mass energy in 2016. All the details about the time profile of luminosity delivered to ATLAS in 2016, 2017, and 2018 runs can be found in~\cite{ATLASluminosity}.}
\label{fig:luminosity_profile_16}
\end{figure}

It should be noted that, in the FLUKA simulations, the delivered luminosity per day throughout an irradiation period was assumed to be constant. However, in reality it varies daily. 
An example of the integrated luminosity delivered and recorded by ATLAS in each day of the 2016 $p$+$p$  run is shown in Fig.~\ref{fig:luminosity_profile_16}. Because the yield of $^{22}$Na correlates to the number of $p$+$p$  collisions and thereby to the integrated luminosity, there is a discrepancy in the estimated activity between a constant and a time-dependent irradiation profile. A correction for the discrepancy has been studied and is shown in Appendix \ref{sec:time_dependent_correction}. The correction was found to be negligible, because the half-life of $^{22}$Na is much longer than the irradiation period. 

%
%
%
%
\section{Activation measurements}
\label{sec:Activation}

As indicated in Sec.~\ref{sec:introduction}, the accuracy of activation calculations by FLUKA has been extensively probed. This applies also to our region of interest \cite{RPTAN,infantino}. Here we aim to provide an additional investigation, looking at the $^{22}$Na activity in BRAN rod samples.  
Because of the cylindrical geometry of the samples, it was's challenging to calculate the acceptance for the activation measurements. To reduce the acceptance error, two independent activation measurements were carried out at UIUC and ANL, 
using a high purity germanium (HPGe) radiation detector and a thallium activated, sodium iodide, NaI(Tl), well-type detector, respectively.
In both cases, the count rate of the $^{22}$Na isotope in each rod sample was measured. 

\subsection{Sample preparations}
\label{sec:BRANcut}
To accurately profile the fused silica's activation the rods were cut into smaller samples, to avoid side-effects due to the activity of neighboring portions of the measured rod. Moreover, cutting the rods into smaller samples was also functional to another analysis not reported in this paper, aimed at measuring the optical transmission of the fused silica rods \cite{Transmission:2022}. Rod 1, 3b, and 6 were chosen to be cut and analyzed because they were the last removed from the prototype and are characterized by higher activities, which provide lower statistical errors, compared to the other rods. 
A length of 10 mm was chosen to meet the requirements of both analyses. To account for the material losses during the cutting, the weights of both the cut sample and the remaining uncut rod were recorded at each iteration. A digital caliper was used to measure the $i^{th}$ segment's maximum length ($L_{i}^{max}$) and minimum length ($L_{i}^{min}$) by rotating the segment 360$^{\circ}$ within the calipers. The remaining rod length, $L_{i}^{rem}$, was also measured using a digital caliper for later calculations. The average cut length for the $i^{th}$ segment, $L_{i}$, was then determined as

\begin{equation}
 L_{i}=\left(\dfrac{L_{i}^{max}+L_{i}^{min}}{2}\right).
  \label{eq:avg_l}
\end{equation}



\subsection{Calibrated volumetric source}
\label{sec:volsource}
To calibrate both detectors, a  $^{22}$Na volumetric source was purchased from Eckert \& Ziegler. The relevant specifications are listed in Tab.~\ref{Tab:volume_source_spec}. The dimensions of the cylindrical source were chosen in order to match those of rod segments produced by cutting the irradiated BRAN rods, as described in Sec.~\ref{sec:BRANcut}. The $^{22}$Na isotope is uniformly distributed throughout the source volume.

\begin{table}[htbp]
\centering
\caption{Specifications for  $^{22}$Na volume source.} 
\begin{tabular}{|l | c | c |}
\hline
Diameter & 10 mm  \\
\hline 
Length   & 10 mm  \\
\hline 
Material & Solid plastic matrix \\
\hline
Density & 1.17 g/cm$^3$ $\pm$ 3\% \\
\hline
Isotope &  $^{22}$Na\\
\hline
Activity & 0.1 $\mu$ Ci $\pm$ 3.3\% \\
\hline
\end{tabular}
\label{Tab:volume_source_spec} 
\end{table}

\subsection{High Purity Germanium (HPGe) Detector (UIUC)}
\label{sec:hpge}
\begin{figure}[htbp]

\includegraphics[width=0.4\textwidth]{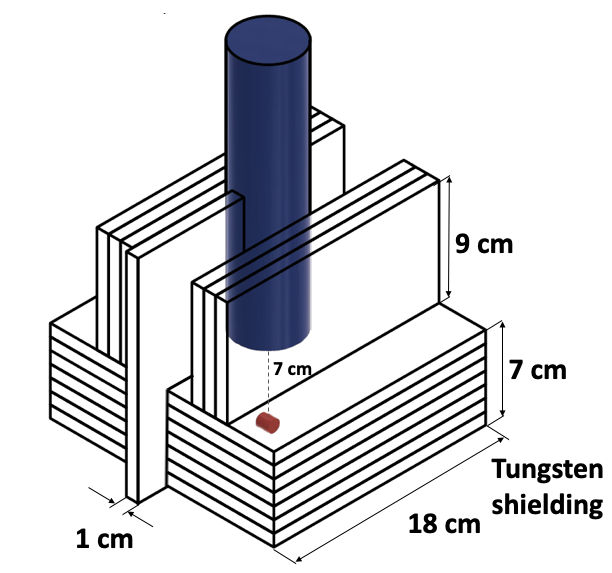}
\caption{Experimental setup of HPGe detector used at the University of Illinois Urbana-Champaign. The transparent plates represent the tungsten shielding. The blue cylinder is the HPGe detector and the red cylinder indicates the location of the sample during measurements.}
\label{fig:HPGe_setup}
\end{figure}

The relative efficiency and energy resolution (FWHM) of the HPGe detector used in this experiment are 19\% and 1.68 keV for the 1.332 MeV peak of $^{60}$Co. A full list of specifications for the HPGe detector is given in Tab.~\ref{Tab:HPGe_specification}. During the measurement, the detector was shielded using tungsten plates to reduce background radiation. The setup at UIUC is shown in Fig.~\ref{fig:HPGe_setup}. Each sample was measured for 15 minutes to achieve a 2\% statistical error for the 1.275 MeV characteristic peak of  $^{22}$Na. For the  $^{22}$Na calibration source, the measurement time was 14 hours, corresponding to a statistical error of 0.1\%. The background was measured for 24 hours and then subtracted from all the measurement results of HPGe presented in this study. 

\begin{table}[htbp]
\centering
\caption{Specifications for the HPGe detector at UIUC.} 
\begin{tabular}{|l | c | c |}
\hline
Model number & ORTEC GEM-10-P4~\cite{ortec_HPGe}  \\
\hline 
Energy resolution   & 1.68 keV @ 1.33 MeV, Co-60  \\
\hline 
Detector diameter & 49.9 mm \\
\hline
Detector length &  54.6 mm \\
\hline
Relative efficiency &  19\% @ 1.33 MeV, Co-60 \\
\hline
\end{tabular}
\label{Tab:HPGe_specification} 
\end{table}

\subsection{Well-type Detector (ANL)}
\label{sec:welltype}
A well-type detector, manufactured by the PerkinElmer Company and with energy resolution $<$10 \% for Cs-137, was used  to measure the  $^{22}$Na activity at ANL. Its specifications are reported in Tab.~\ref{Tab:Well_det_specification}. The activity of each sample was obtained from the built-in analysis software (PerkinElmer Wizard$^2$ 2480\cite{wizard2480}), including the uncertainty on the count rate.
Each sample was measured for a time sufficient to reach a statistical error less than 1\% for the 1.275 MeV characteristic peak of $^{22}$Na. The volumetric  $^{22}$Na calibration source was measured for 1 hour, achieving 0.39\% statistical error.
  
 \begin{table}[!htbp]
\centering
\caption{Specifications of the well-type detector at ANL.} 
\begin{tabular}{|l | c | c |}
\hline
Model number & 2480 Wizard gamma counters  \\
\hline 
Material & Na(Tl) crystal  \\
\hline 
Energy resolution   &  $<$ 10 \% for Cs-137  \\
\hline 
Crystal diameter & 75 mm \\
\hline
Crystal height &  80 mm \\
\hline
Relative efficiency &  47\% for Cs-137 \\
\hline
\end{tabular}
\label{Tab:Well_det_specification} 
\end{table}

%
%
%
%
\section{Data analysis}
\label{sec:DataAnalysis}

All the relevant aspects of the data analysis for the activity measurements are presented in this section. Details on the gamma count rate analysis for UIUC measurements can be found in Sec.~\ref{sec:UIUC_analysis}. Sec.~\ref{sec:estimate_Na_activity} describes how the $^{22}$Na activity of BRAN rod samples was estimated from the measured count rate using the calibration source. Sec.~\ref{sec:material_loss} and Sec.~\ref{sec:cooling_correction} discuss the material losses and cooling corrections, respectively.


\subsection{Estimation of count rate for HPGe detector} 
\label{sec:UIUC_analysis}
The  $^{22}$Na count rate was estimated from the spectrum measured with the HPGe detector. The 1.275 MeV gamma peak of  $^{22}$Na was targeted because its branching ratio is 99.9\%. After subtracting the background and the baseline from the targeted peak, a Gaussian fit was applied to estimate the count rate, defined as the area within the full width tenth maximum (FWTM) of the Gaussian fit.  


\subsection{Estimation of  $^{22}$Na activity}
\label{sec:estimate_Na_activity}
The acceptance of the experimental setup must be taken into account when estimating the activity of a volumetric isotope's sample.
Two types of acceptance effects are relevant for the measurement presented in this paper. The first is introduced by differences between the geometry of the  calibration source and the fused silica samples. To account for this effect, the volumetric source, described in Sec.~\ref{sec:volsource}, was used to calibrate the setup. Because the geometry of the calibrated source and the BRAN samples are consistent with one another, as well as the positioning of samples within the experimental setup, the acceptance was taken to be equivalent. This conclusion relies on the assumption that the activity of both the samples and the calibrated source are homogeneous. 
The second effect, introduced by the intrinsic geometrical acceptance of the experimental setup for a single photon measurement, can be cancelled out by taking the ratio of the count rates recorded for the sample and calibration source. 

The activity of the  $i^{th}$ sample is calculated as
\begin{equation}
\label{eq:activit_estimation}
A_{i} = \frac{I_{i}} {I_{CS}} \cdot A_{CS},
\end{equation}
where $I_{i}$ and $I_{CS}$ are the measured count rate of the $i^{th}$ sample and the volumetric calibration source, respectively, while $A_{CS}$ is the known activity of the calibration source. The relative uncertainty on $A_{i}$, $R_{A_{i}}$, was obtained from the error propagation of each component in Eq.~\ref{eq:activit_estimation} and is given as
\begin{equation}
\label{eq:uncertainty_estimation}
R_{A_{i}} = \sqrt{R_{I_{i}}^2 + R_{I_{CS}}^2 + R_{A_{CS}}^2 },
\end{equation}
where $R_{X}, X \in \{I_i, I_{CS}, A_{CS} \}$ represents the relative error of the variables entering Eq.~\ref{eq:activit_estimation}.


\subsection{Material losses correction}
\label{sec:material_loss}
A correction was applied to account for the decrease in activity of each sample due to material losses during the cutting process. The correction coefficient, $C_{i}$,  was based on the weight of lost material per sample: 
\begin{equation}
\label{eq:material_loss}
C_{i} = \frac{W_{\circ}}{W_{i}}, 
\end{equation}
where $W_{\circ}$ is the weight of a precisely 10 mm long sample of BRAN rod and $W_{i}$ is the measured weight of the $i^{th}$ sample. The activity of each sample was corrected by applying the correction coefficient
\begin{equation}
A^{*}_{i} = C_i \cdot A_{i}.
\end{equation}
It is assumed that the activity is homogeneous within each sample.

In order to properly map the activity results in the FLUKA simulation, the center of the samples within the BRAN rod prior to cutting was calculated as 
\begin{equation}
P_{i} = L_{rem,i} + L_{loss} + 0.5 \cdot L_{i}, 
\end{equation}
where $P_{i}$ is the center of the $i^{th}$ sample in the uncut rod, $L_{rem,i}$ is the length of the remaining uncut rod after the $i^{th}$ cut and $L_{i}$ is the length of the $i^{th}$ sample, calculated using Eq.~\ref{eq:avg_l}. The amount of rod lost due to cutting each side, $L_{loss}$, is assumed to be 0.38 mm, equal to twice the width of the saw blade\footnote{Please note that, for the first cut of the rod, $L_{loss}$ is assumed to be 0.19 mm since those segments are obtained with only one cut.}. $P_{i}$ was calculated immediately after each cut, allowing for an accurate measure of $L_{i}^{rem}$.


\subsection{Cooling correction}
\label{sec:cooling_correction}

To allow for a comparison of activity estimates calculated for a given segment, as well as comparison between the measurements and the FLUKA simulations, the activity for all samples was normalized to the same reference date (12/14/2019). The isotope activity for each segment was estimated making use of UIUC and ANL measurements separately. 

The Beer–Lambert law~\cite{lamarsh2001introduction} is used to extrapolate the activity to the reference date:
\begin{equation}
 A_{t} = A_{0}e^{-\lambda t},
 \label{eq:Beer–Lambert}
\end{equation}
where $A_{t}$ is the activity on the reference date, $A_{0}$ is the activity on the measurement date, $t$ is the time between the measurement and reference date, and $\lambda$ is the decay constant of the isotope being measured. For the  $^{22}$Na isotope considered in this study, the decay constant is 0.2664~yr$^{-1}$, derived from half-life of 2.6 year~\cite{Basunia:2015qqx}.

%
%
%
%
\section{Results and discussion}
\label{sec:Results}
\subsection{Experimental results}
\begin{figure}[!htbp]
\includegraphics[width=0.44
\textwidth]{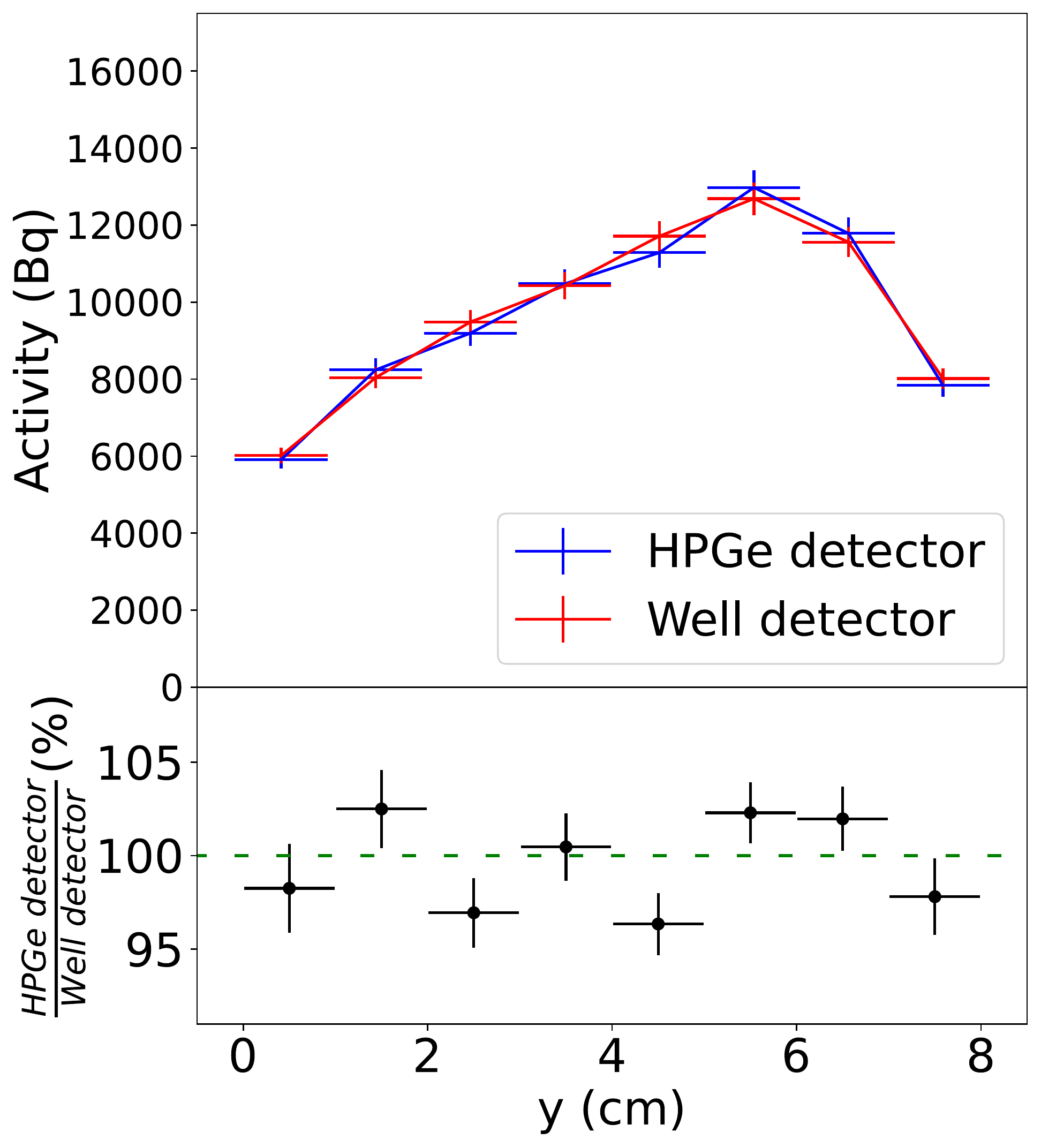}
\caption{Comparison of Rod 1 activity measurements performed at UIUC, using a HPGe detector (Sec.~\ref{sec:hpge}), and ANL, using a well-type detector (Sec.~\ref{sec:welltype}). The errors on the abscissa represent the half-length of the samples. The bottom panel shows the ratio between the two measurements. A dashed green line that corresponds to unity is drawn for direct comparison. Please note that there is an additional 3.3\% correlated uncertainty on the calibration source activity that is not included in the ratio uncertainties. Results have been normalized to a reference date of 12/14/2019. }
\label{fig:count_comparison_rod1}
\end{figure}

\begin{figure}[!htbp]
\includegraphics[width=0.44
\textwidth]{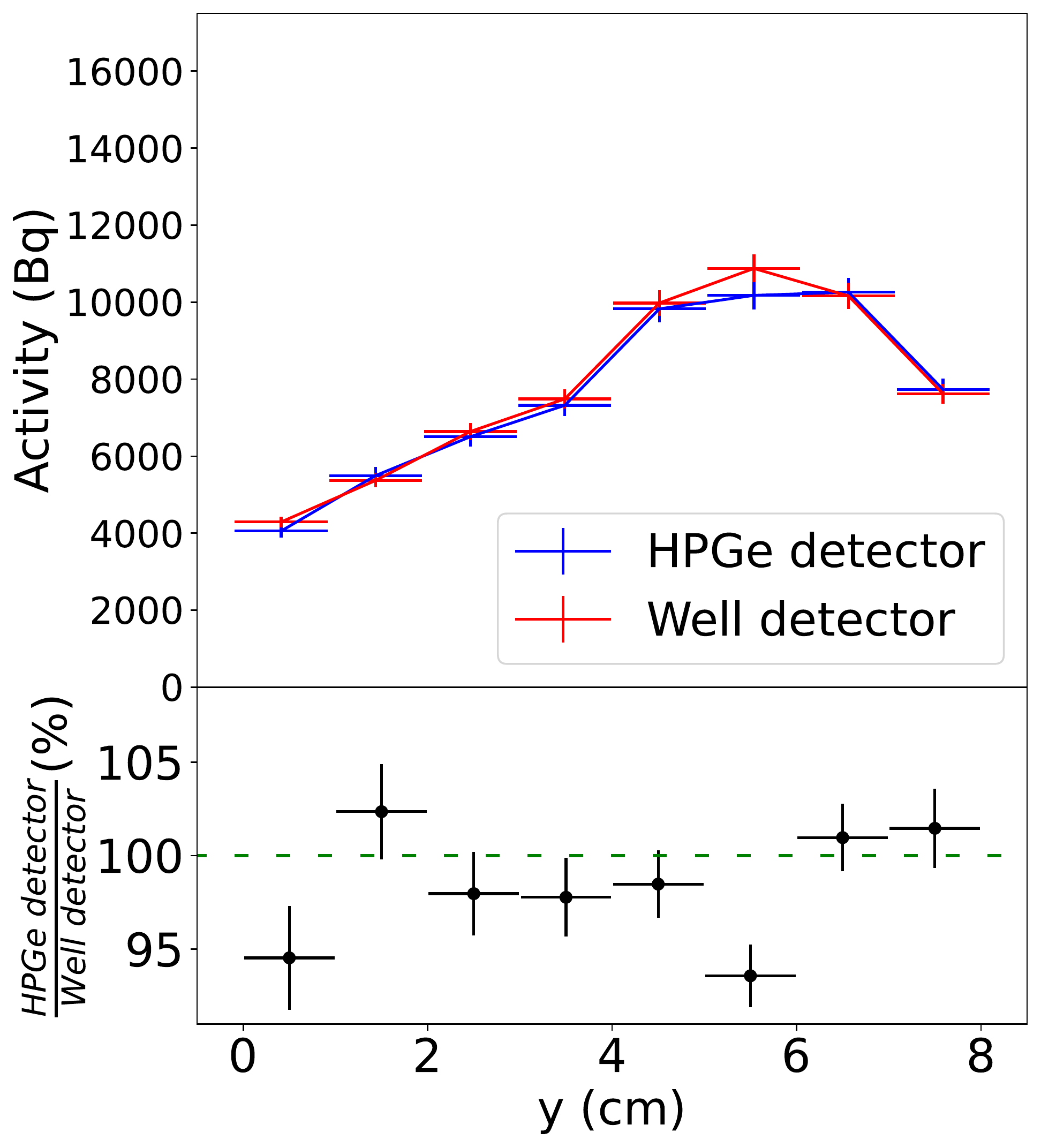}
\caption{Comparison of Rod 3b activity measurements using a HPGe and a well-type detector. Refer to the caption of Fig.~\ref{fig:count_comparison_rod1} for details related to the plotting.}
\label{fig:count_comparison_rod3b}
\end{figure}

\begin{figure}[!htbp]
\includegraphics[width=0.44\textwidth]{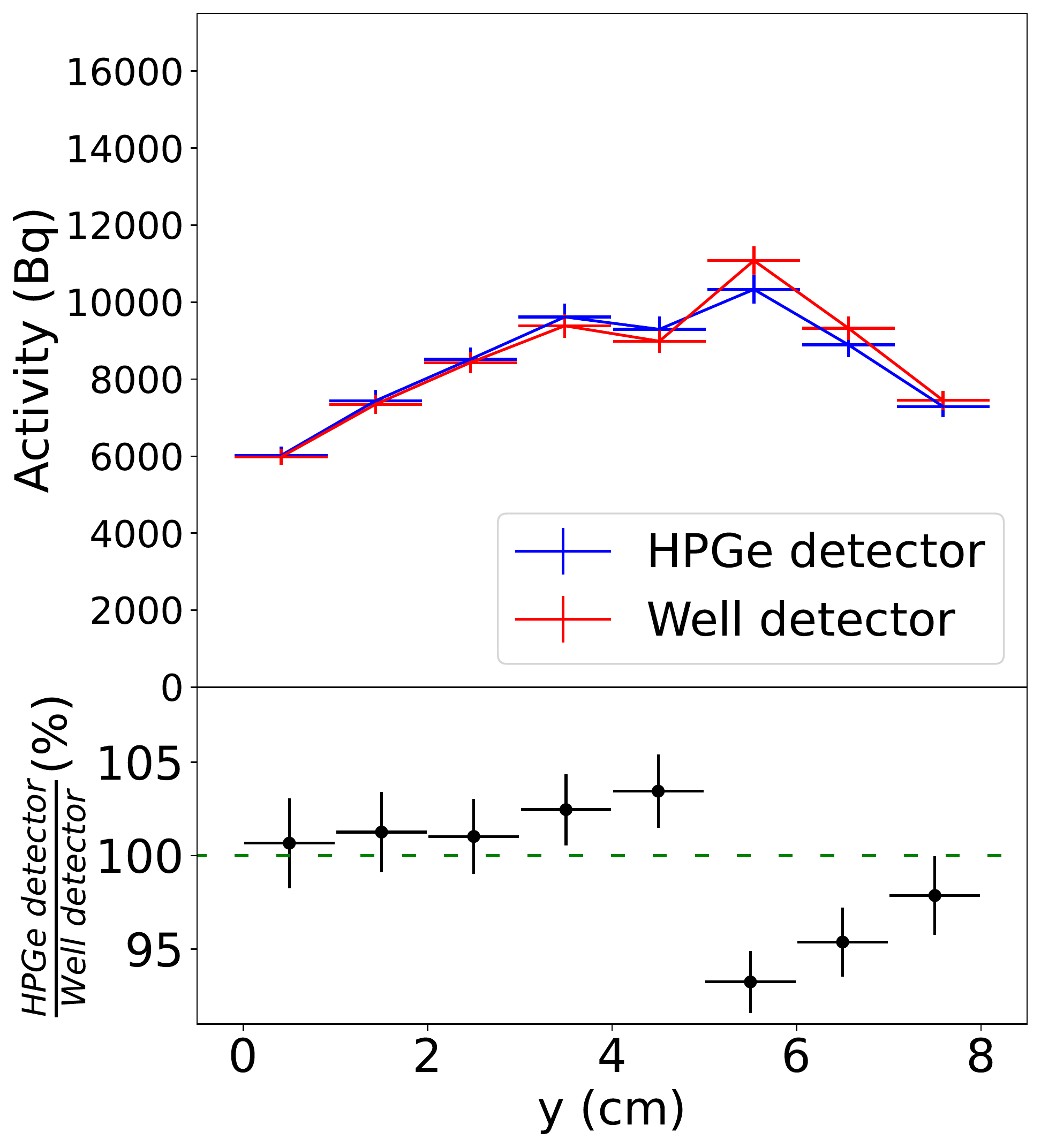}
\caption{Comparison of Rod 6 activity measurements using a HPGe and a well-type detector. Refer to the caption of Fig.~\ref{fig:count_comparison_rod1} for details related to the plotting. }
\label{fig:count_comparison_rod6}
\end{figure}

The comparisons between the  $^{22}$Na activity measurements done at UIUC (HPGe) and ANL (well-type detector) for Rods 1, 3b and 6 are shown in Fig. \ref{fig:count_comparison_rod1}, \ref{fig:count_comparison_rod3b}, and \ref{fig:count_comparison_rod6}, respectively. At this stage, both results are not corrected for material losses. The blue and red curves represent the activity of each segment measured with the HPGe and well-type detector, respectively. Due to the crossing angle, the peak of the activity is located around 5 cm from the bottom of the rod. The lower portion of the figures shows the ratio between the two measurements, that demonstrates consistency between the two curves in all the cases. 

\subsection{Comparisons between experimental and simulation results}

\begin{figure}[!htbp]
\includegraphics[width=0.45\textwidth]{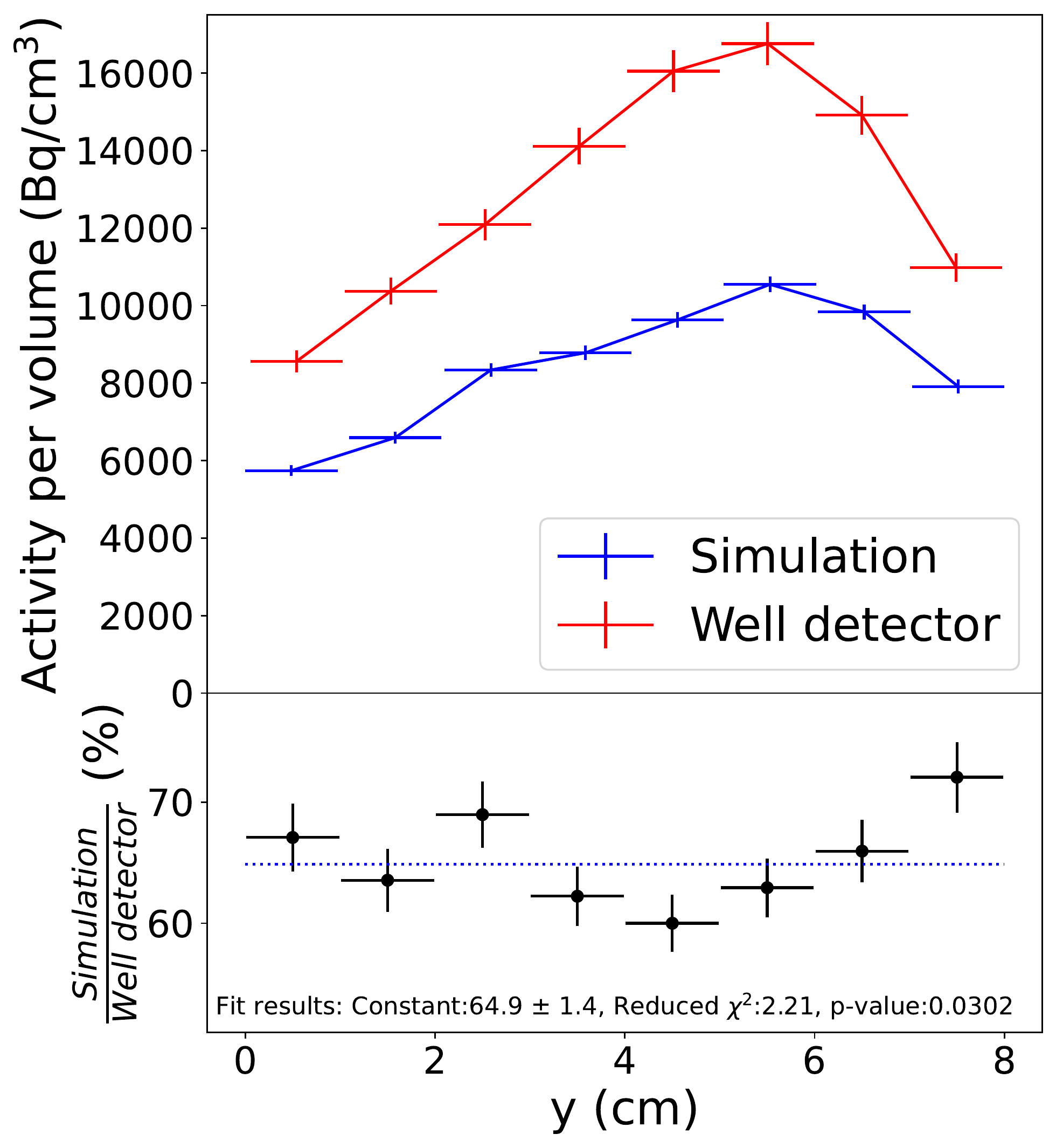}
\caption{Comparison of activity measurements at ANL, using a well-type detector, and FLUKA simulations for Rod 1. Results have been normalized to a reference date of 12/14/2019. The dotted line represents the fit of the ratio between data and simulations.}
\label{fig:sim_exp_rod1}
\end{figure}

\begin{figure}[!htbp]
\includegraphics[width=0.45\textwidth]{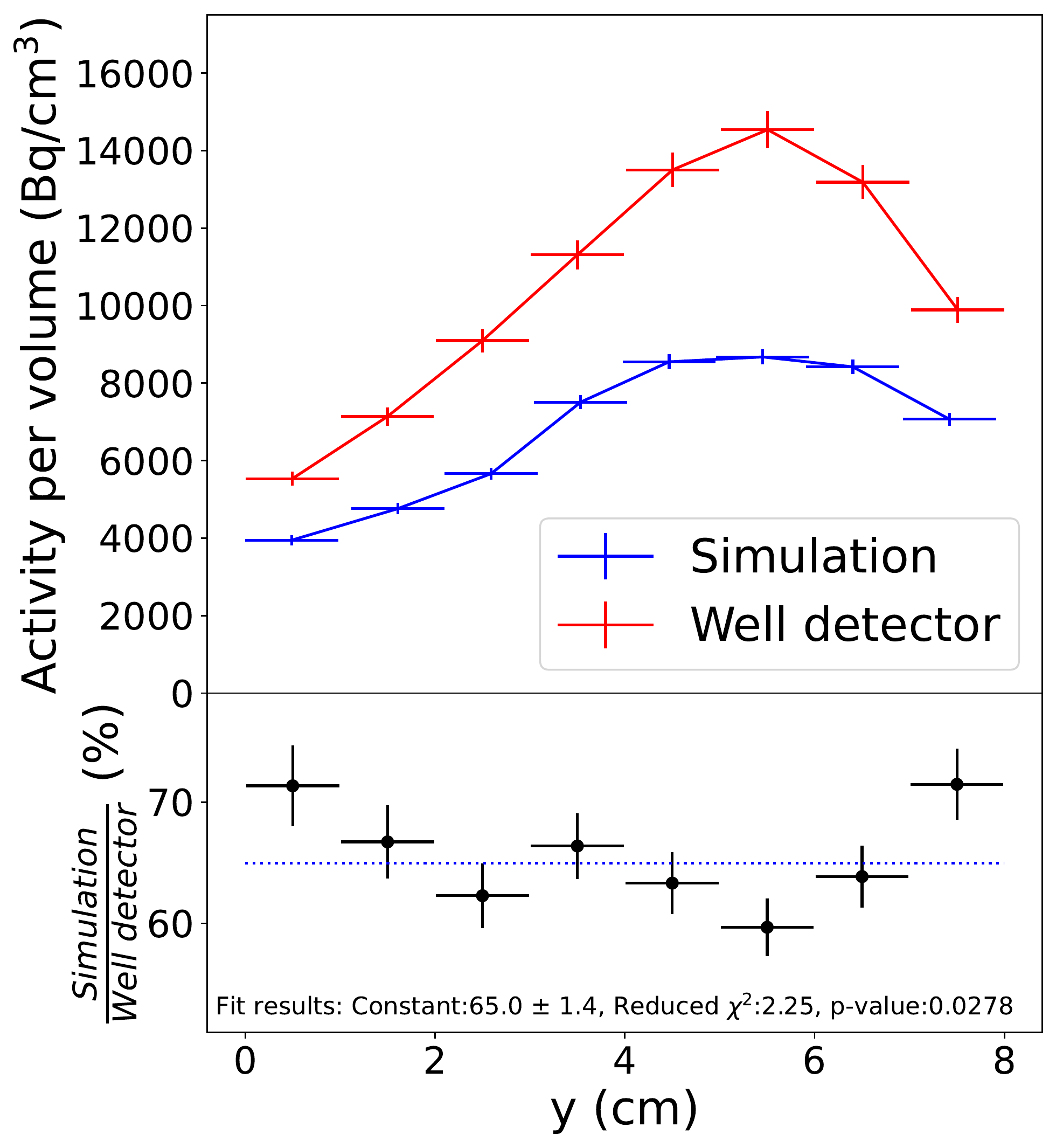}
\caption{Comparison of activity measurements and FLUKA simulations for Rod 3b. Refer to the caption of Fig.~\ref{fig:sim_exp_rod1} for details related to the plotting. }
\label{fig:sim_exp_rod3b}
\end{figure}

\begin{figure}[!htbp]
\includegraphics[width=0.45\textwidth]{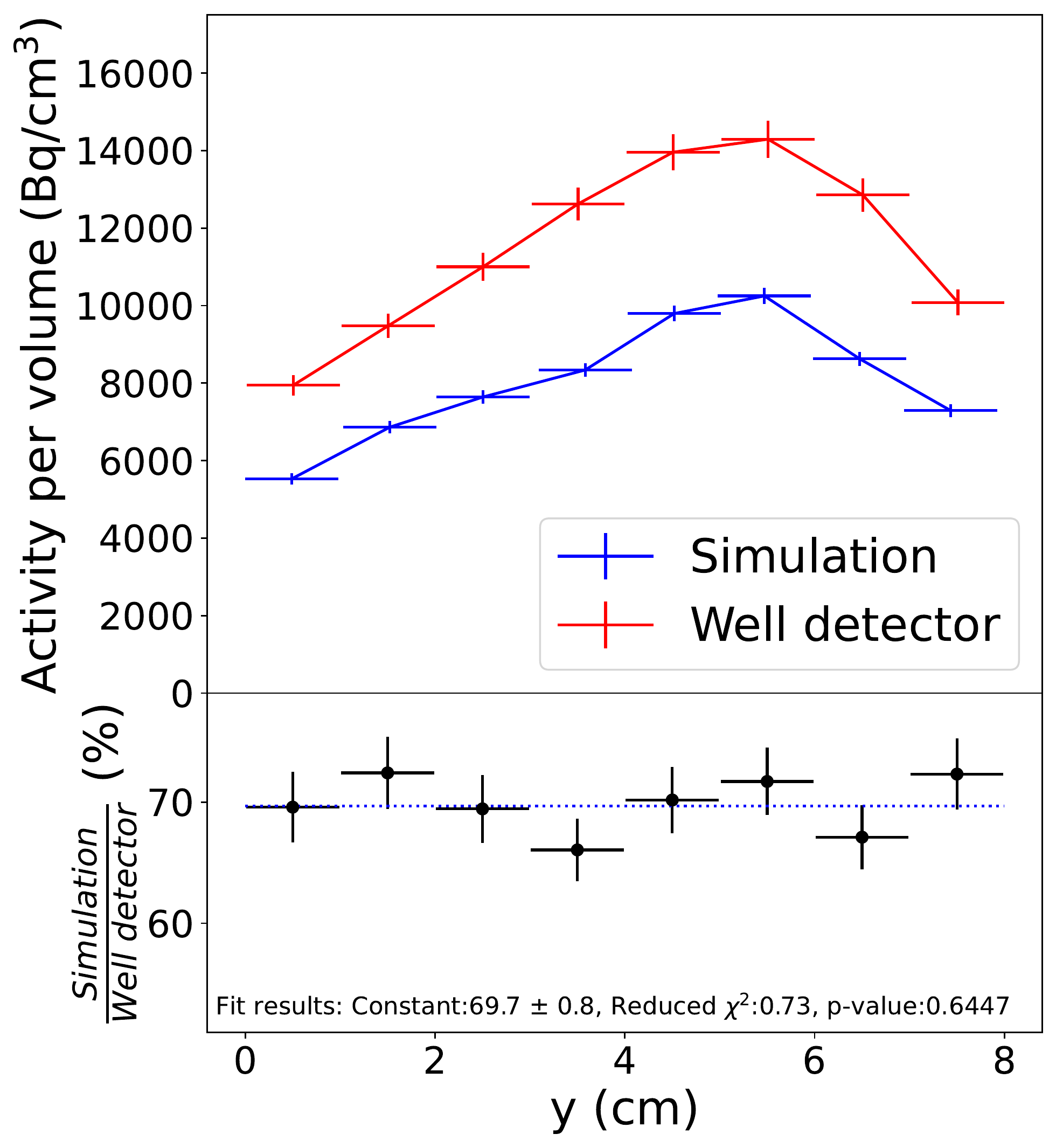}
\caption{Comparison of activity measurements and FLUKA simulations for Rod 6. Refer to the caption of Fig.~\ref{fig:sim_exp_rod1} for details related to the plotting.}
\label{fig:sim_exp_rod6}
\end{figure}

Fig. \ref{fig:sim_exp_rod1}, \ref{fig:sim_exp_rod3b}, and \ref{fig:sim_exp_rod6} show the specific activity comparison between the measurement and simulation for Rod 1, 3b, and 6, respectively. The blue curve represents the data collected using the well-type detector, while the red curve corresponds to the simulation result obtained from FLUKA. The experimental results are corrected for material losses (Sec.~\ref{sec:material_loss}). Both experimental and simulation results are normalized to the same reference date, as described in Sec.~\ref{sec:cooling_correction}. The relative statistical uncertainty for simulation and experimental results is 3\% and 4\%, respectively. The FLUKA simulation reproduces the location of the activity peak observed in the experimental measurements around 5 cm. On the other hand, it underestimates the activity of $^{22}$Na by 30-35 \% compared to the experimental results. The $^{22}$Na production is due to nuclear reactions in silicon that, according to simulations, are mostly induced by neutrons (40\%) and charged pions (45\%), with the rest coming mainly from protons. Based on Figs.~\ref{fig:22Nabyp} and \ref{fig:22Nabyn}, such a discrepancy may hint at an underestimation of the $^{28}$Si($\pi^{\pm}$,$\star$)$^{22}$Na cross sections in FLUKA, for which no direct benchmarking data was found. If one assumes that the missing 30-35 \% is exclusively due to the charged pion cross section estimate, this should be increased up to a factor of two.  

%
%
%
%
\section{Conclusions}
\label{sec:Conclusions}

The BRAN prototype was installed in the ATLAS TAN, one of the highest radiation locations of the LHC, in March 2016. Fused silica samples inserted in the prototype were irradiated during $p$+$p$  runs from 2016 to 2018. The irradiation of the fused silica caused nuclear reactions, producing radioactive isotopes within the rods. 
By measuring the isotope activity, it is possible to benchmark FLUKA's performance in describing material activation in a complex radiation environment. In this paper, the experimental and simulation results of the $^{22}$Na activity in the rods were presented.

To profile the $^{22}$Na isotope activity, the rods were cut into 1 cm samples. The cut samples were measured by two independent setups to reduce the acceptance error of the measurements. A $^{22}$Na cylindrical calibration source with a 3.3\% uncertainty was used to estimate the activity of both measurements. The activity results were normalized to the same reference date for comparison. A constant function was fitted to the ratio between the two independent measurements and demonstrated their agreement.

Dedicated FLUKA simulations of 13 TeV center-of-mass $p$+$p$ inelastic interactions in IP1 were performed to calculate the absorbed dose map across the TAN, the BLM dose pattern along the beam line, as well as the residual activity distribution in the fused silica rods, exploring at the same time the $^{22}$Na origin.

The experimental activity results were corrected for material losses experienced while cutting the samples and compared to simulation results. The $^{22}$Na activity estimated by FLUKA simulations turned out to be 30-35 \% lower than the experimental measurements. This level of accuracy, referring to one specific isotope, represents a rather satisfactory agreement, considering the complexity of the radiation field originated by the primary beam collisions and the accelerator layout. In parallel, an excellent reproduction of the BLM measurements over the considered 130 m long tunnel section was achieved.

A new irradiation campaign of Heraeus fused silica samples will be carried out in the TAN during Run 3. Thanks to the higher luminosity delivered by the accelerator, and the much closer position of the samples to the shower maximum, the accumulated dose in the samples will surpass the one presented in this paper by at least one order of magnitude. 
The fused silica samples are 1 cm long cylinders and polished on both ends. This will serve to reduce geometrical variations, prevent the need for cutting radioactive material, and minimize preparations required for activation and transmission measurements. These samples will enable the possibility of measuring other isotopes with shorter half-lives (e.g. $^7$Be), thereby providing further benchmarks for FLUKA's predictions. 

\vspace{0.4cm}
\section*{Acknowledgements}
This work is in part supported by the National Science Foundation, Grants no. PHY-1812377, PHY-1812325 and PHY-2111046. 
We would like to thank Prof. Angela Di Fulvio of the Nuclear, Plasma, and Radiological Department at UIUC for allowing us to carry out measurements using the HPGe detector in her laboratory.


\bibliography{PRBA_main}

\begin{appendices}

\section{Activation correction for constant luminosity in simulation}
\label{sec:time_dependent_correction}

A correction for the scored activation in the simulations with a constant luminosity was studied. The yield, $N_0$, of $^{22}$Na per day at time $t_0$ is assumed to be dependent on the delivered luminosity per day, $\mathcal{L}_0$. By 
\begin{equation}
  \label{eq:decay}
  A_0 = \lambda N_0,
\end{equation}
where $A_0$ is the activity of $^{22}$Na and $\lambda$ is a decay constant of the $^{22}$Na, the activity, $A_0$, is proportional to the integrated luminosity per day, $\mathcal{L}_0$.

To include the decay factor of $^{22}$Na, we define a equivalent luminosity, $\mathrm{L}_{t}$, as 
\begin{equation}
 \mathrm{L}_t = \mathcal{L}_0 \cdot e^{-\lambda (t - t_0)},
\end{equation}
where $t$ is the time of the measurement. The equivalent luminosity is proportional to the activity measured at time $t$.

The total activity generated from a LHC luminosity profile is proportional to the superposition of the equivalent luminosity, $\mathrm{L} (y)$, in a given year $y$ and described as 

\begin{equation}
  \label{eq:proptional_constant}
    \mathrm{L(y)} = \sum\limits_{t =t_{0}}^{t_{1}}\mathcal{L}_t (y) \cdot e^{ -\lambda (t - t_{1}) }
\end{equation}
where $\mathcal{L}_t$ is integrated luminosity of a given day $t$ in a given year $y$, $t_0$, and $t_1$ are the the first and last date of the irradiation period, respectively. For a constant luminosity, Eq~\ref{eq:proptional_constant} could be simplified to 

\begin{equation}
    \mathrm{L}_c (y) = \bar{\mathcal{L}} (y)\sum\limits_{t =t_{0}}^{t_{1}} e^{ -\lambda (t - t_{1}) }, 
\end{equation}
where $\bar{\mathcal{L}}$ is the average luminosity in a irradiation period in a given year.

The corrected activity of the $^{22}$Na generated from a luminosity profile at LHC, $A_{LHC}(y)$, at a given year $y$ could be obtained from the activity with constant luminosity $A_{sim} (y)$, as: 

\begin{align}
  \label{eq:Exp_correction}
 A_{LHC} (y) &= A_{sim} (y) \cdot C (y)
\end{align}
where
\begin{align}
C(y) &= \frac{\mathrm{L}_c (y)}{\mathrm{L} (y)}
\end{align}

Based on the luminosity profiles and the total luminosity from~\cite{ATLASluminosity} in Run 2, C(y) is 1.02, 0.99, and 0.99 for 2016, 2017, and 2018 respectively. The correction factors are small because the half-life (2.6 yrs) of $^{22}$Na is much longer than the irradiation period (i.e. 198 days).

\end{appendices}

\end{document}